\documentclass[preprint2]{aastex}

\begin{document}
 
\title{A Primordial Origin for Misalignments Between Stellar Spin Axes and Planetary Orbits}  
\author{Konstantin Batygin} 

\affil{Division of Geological and Planetary Sciences, California Institute of Technology, 1200 E. California Blvd., Pasadena, CA 91125}
\affil{Institute for Theory and Computation, Harvard-Smithsonian Center for Astrophysics, 60 Garden St., Cambridge, MA 02138} 
\email{kbatygin@gps.caltech.edu}
 
\maketitle

\textbf{\\ The presence of gaseous giant planets whose orbits lie in extreme proximity to their host stars (``hot Jupiters"), can largely be accounted for by planetary migration, associated with viscous evolution of proto-planetary nebulae$^1$. Recently, observations of the Rossiter-McLaughlin effect$^2$ during planetary transits have revealed that a considerable fraction of detected hot Jupiters reside on orbits that are misaligned with respect to the spin-axes of their host stars$^3$. This observational fact has cast significant doubts on the importance of disk-driven migration as a mechanism for production of hot Jupiters, thereby reestablishing the origins of close-in planetary orbits as an open question. Here we show that misaligned orbits can be a natural consequence of disk migration. Our argument rests on an enhanced abundance of binary stellar companions in star formation environments, whose orbital plane is uncorrelated with the spin axes of the individual stars$^{4,5,6}$. We analyze the dynamical evolution of idealized proto-planetary disks under perturbations from massive distant bodies and demonstrate that the resulting gravitational torques act to misalign the orbital planes of the disks relative to the spin poles of their host stars. As a result, we predict that in the absence of strong disk-host star angular momentum coupling or sufficient dissipation that acts to realign the stellar spin axis and the planetary orbits, the fraction of planetary systems (including systems of hot Neptunes and Super-Earths), whose angular momentum vectors are misaligned with respect to their host-stars should be commensurate with the rate of primordial stellar multiplicity.}

\begin{figure}[t]
\includegraphics[width=1\columnwidth]{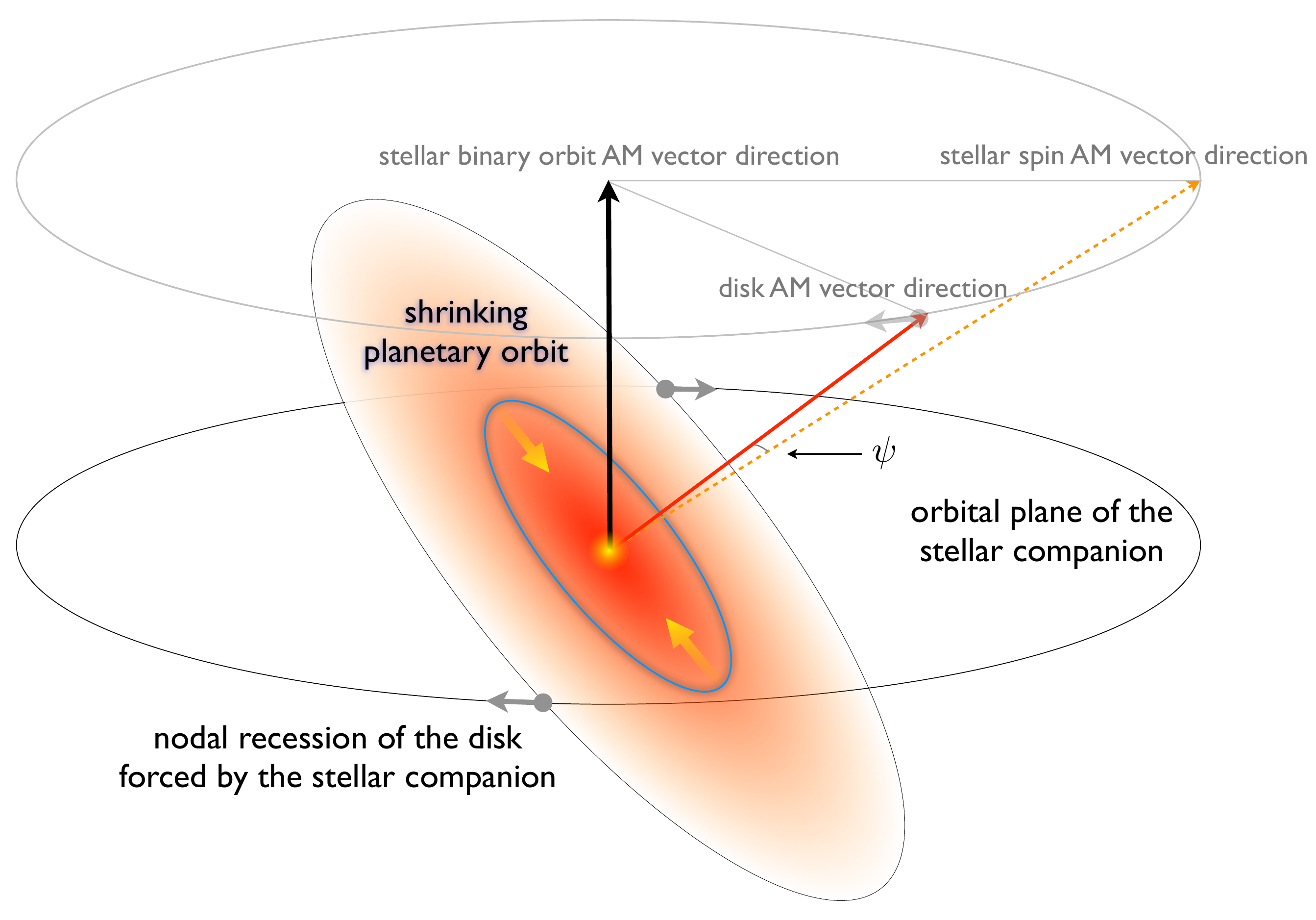}
\caption{\small{Geometrical setup of the problem. This figure depicts a schematic representation of the production of misaligned close-in planets through disk-driven migration in binary systems. The adiabatic response of a self-gravitating disk to long-term perturbations by a stellar companion is the recession of its ascending node, as defined by the orbital plane of the stellar companion. The recession of the disk's angular momentum (AM) vector about the stellar binary orbital AM vector appears to be an excitation of misalignment between the stellar spin-axis and the disk, $\psi$, in the star's reference frame.}}
\end{figure}	

The obliquities (angles between the planetary orbits and the stellar spins) of detected planetary orbits range from almost perfectly aligned prograde to almost perfectly aligned retrograde systems$^7$. Previously, the misalignment between planetary orbits and stellar spin axes had been attributed to post-nebular multi-body interactions. Most notably, Kozai cycles with tidal friction$^{8,9,10}$, planet-planet scattering$^{11,12}$, and chaotic secular excursions$^{13}$ have been invoked as a means of producing misaligned planets. These mechanisms are likely responsible for a few specific examples (e.g. the extreme eccentricity of HD80606b is almost certainly due to Kozai resonance with the stellar companion HD80607$^{8}$), however it is unlikely that they can explain misaligned hot Jupiters as a population. For example, the Kozai mechanism can be stifled by forced apsidal precession in multi-planet systems$^{13}$. Likewise, within the context of planet-planet scattering and secular chaos, the allowed parameter range is limited, since the production of close-in orbits requires the timescale for tidal capture to be considerably shorter than that for eccentricity growth$^{12}$, while demanding the associated tidal heating to be sufficiently small to not over-inflate the planet beyond its Roche-lobe$^{14}$. Additionally, the observed presence of mean-motion resonances among giant planets on wide orbits (which rely on smooth, convergent migration to congregate$^{15}$), provides further motivation for the development of a unified model for disk migration that is capable of producing misaligned orbits.

The dynamics of self-gravitating proto-planetary disks under external perturbations can be extremely complex, making precise quantitative modeling computationally unfeasible. Consequently, here we concentrate on characterization of the qualitative physical behavior of the system and utilize classical perturbation methods to obtain a solution. In the spirit of secular theory$^{16}$, we model the proto-planetary disk as a series of initially planar, circular, concentric massive wires that interact gravitationally. Our model is based on the Gaussian averaging method$^{17,18}$ and the gravitational potential is softened in order to partially account for the discrete representation of the disk. The effects of dissipative fluid forces within the disk are neglected. The perturbing body is also modeled as a massive ring, but is eccentric ($e'=0.5$) and inclined with respect to the disk by an inclination $i'$. A detailed description of the model and its inherent assumptions is presented in ref.[$19$] while the particularities of our implementation are stated in the Supplementary Information (SI). 

\begin{figure}
\includegraphics[width=1\columnwidth]{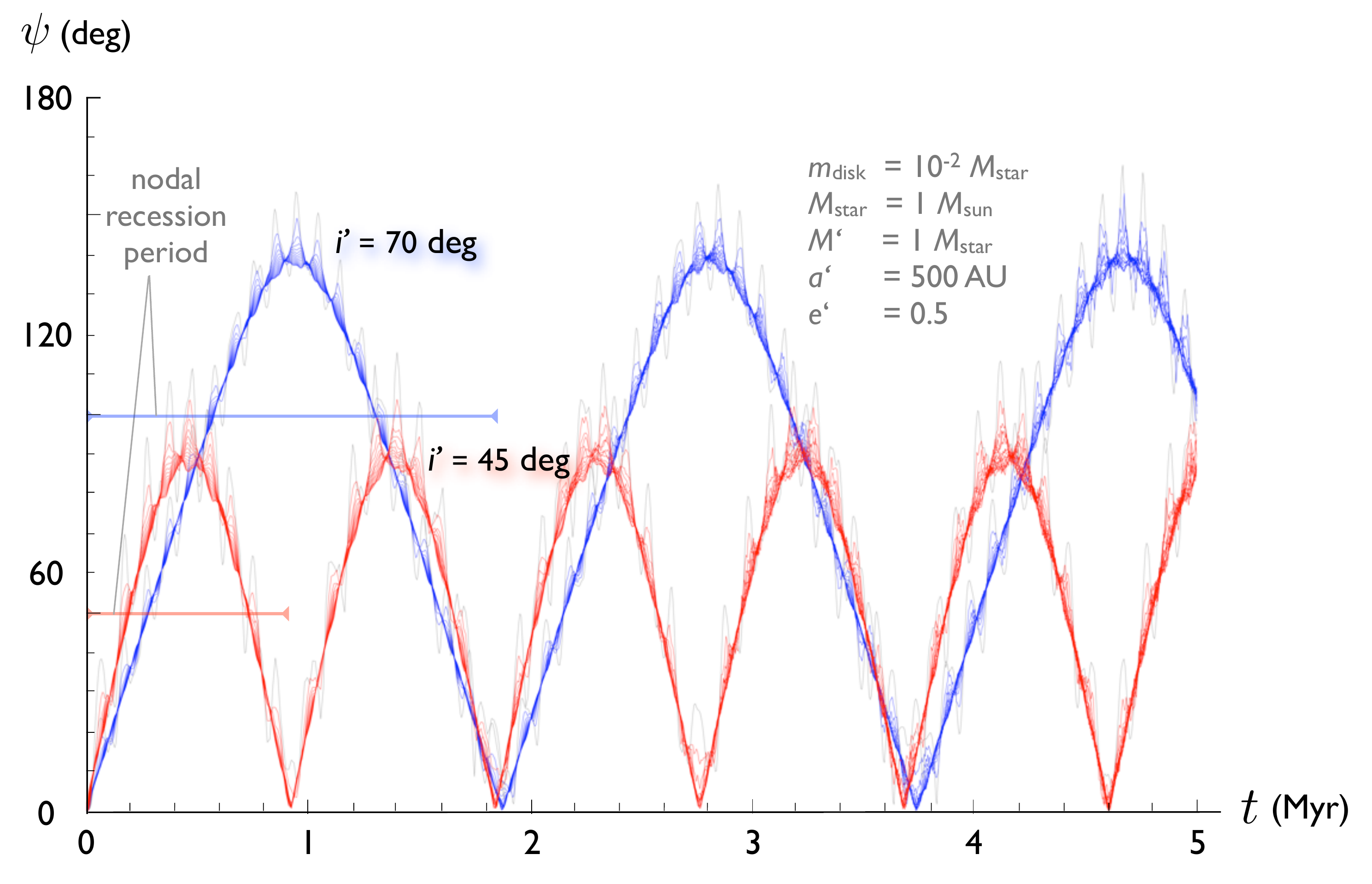}
\caption{\small{Excitation of disk-star misalignment. The time evolution of the misalignment angle, $\psi$ is shown. The considered $m_{disk} = 10^{-2} M_{sun} \simeq 10 M_{Jup}$ nebula has a surface density profile of the form $\Sigma \propto r^{-1}$, extends between $a_{in} = 1$ AU and $a_{out} = 50$ AU, and orbits a $M_{star} = 1M_{sun}$ host star. The plotted curves represent the dynamical states of the disk annuli, where every other ring is plotted. The gray curves depict the inner-most annuli. The close proximity of the rings to to each-other demonstrates that the disk remains locally un-warped and acts as a rigid body to a good approximation. The extent of rigidity is largely controlled by the disk mass, with heavier disks remaining closer to their mid-plane. The host star is assumed to be a slow rotator and its spin pole is held fixed for simplicity. The perturbing $M' = 1M_{sun}$ binary companion lies $a'=500$AU away, has an eccentricity of $e' = 0.5$, and is inclined with respect to the disk by $i'=45^{\circ}$ (red) and by by $i'=70^{\circ}$ (blue). Throughout the duration of the integration, the annuli of the disk never attain significant eccentricity ($e \lesssim 0.1$ - see SI for an in-depth discussion). This calculation was performed using a conservative softened Gaussian averaging model (with $N_{\rm{rings}} = 31$ - see SI), and thus contains no restrictions on the secular dynamics of the system, but ignores dissipative forces of the gas. The nodal recession periods characteristic of this setup are $\mathcal{T}_{disk} \simeq 0.9$Myr ($i'=45^{\circ}$) and $\mathcal{T}_{disk} \simeq 1.8$Myr ($i'=70^{\circ}$). The results in this figure can be translated to other system parameters by noting that the the maximum misalignment attainable by the disk is roughly twice the disk/binary orbit inclination and that the recession period scales approximately as $\mathcal{T}_{disk} \propto a'^3 / (M'  \cos i' (1 + 3 e'^2 /2))$ (see also Figure 3).}}
\end{figure}	

A self-gravitating disk will preserve an untwisted structure and act as a rigid body, provided that the characteristic timescale of the external perturbation greatly exceeds that of the disk's self-interaction$^{20}$. Mathematically, this amounts to a statement of adiabatic invariance of the phase-space area occupied by a single secular cycle within the disk$^{21}$. If this condition is satisfied, the external perturber's sole effect is to induce a recession (i.e. drift) of the ascending node of the disk, as defined by the plane of the stellar orbit. The embedded planetary orbit will also adiabatically follow the disk. 

In the reference frame of the host star, the nodal recession of the disk will appear as a cyclic excitation of inclination between the disk and the stellar spin axis (see Figure 1), provided that the host star's angular momentum vector does not adiabatically trail the disk. For this to hold true, the characteristic interaction timescale between the disk and the stellar spin-axis, $\mathcal{T}_{star}$ must exceed the disk's nodal recession timescale, $\mathcal{T}_{disk}$, by a considerable amount (i.e. angular momentum coupling between the disk and the host star must be non-adiabatic)$^{9}$. The former can be estimated by modeling the stellar rotational bulge as an inertially equivalent orbiting ring, effectively reducing the characteristic interaction timescale to the forced nodal recession period of the ring .

Observations suggest that rotational periods of T Tauri stars, whose masses exceed $M>0.25 M_{\odot}$, form a bimodal distribution where fast and slow rotators are centered around $\sim 2$ days and $\sim 8$ days respectively, with a preference for slow rotation at higher masses$^{22}$. Thus, for typical pre-main-sequence stars, we obtain $\mathcal{T}_{star} \sim 10$ Myr and $\mathcal{T}_{star} \sim 0.3$ Myr for slow and fast rotators respectively (see SI for details). As will be shown below, this suggests that the adiabatic trailing of the stellar spin axes will only prevent excitation of mutual misalignment for fast rotators. Furthermore, the disk-star angular momentum transfer will likely be unimportant in mature disks because of low accretion rates$^{23}$.

In addition to avoiding the adiabatic trailing of the host stars, the prominence of the mechanism described here is determined by the abundance and longevity of wide stellar binary systems in star formation environments, since the misalignment angle, $\psi$, becomes frozen in when the binary companion is stripped away or when the proto-planetary disk dissipates (nevertheless, a long-lived binary companion can act to misalign a mature planetary system with its host star$^{24}$). While it is tempting to estimate the frequency with which significant misalignments are attained by this mechanism via population-synthesis, the enormous range and vast observational uncertainties in the input parameters would render such a calculation of little practical use. In particular, although observations of the Taurus-Auriga star-forming region$^{25}$ suggest that the orbital distribution of young solar-type binaries is roughly log-flat with an overall binary fraction of $\sim 40\%$, it is noteworthy that the process of wide binary formation also appears to exhibit environmental dependence$^{5}$. Simultaneously, the rate at which wide binaries get disrupted in birth clusters depends sensitively on the local densities within the clusters$^{6}$, which remain observationally elusive, as the majority of stars are born in aggregates that dissolve quickly (on timescales of a $\sim \rm{few} \times 10$ Myr or less)$^{26}$. Still, the above conditions likely imply that the timescale for excitation of significant misalignment should be considerably less than $\sim 10$Myr.

In systems where self-gravity is strong enough to maintain the effective rigidity of the disk, fast circulation of the disk's argument of perihelion also ensures adiabatic eccentricity dynamics. Specifically, this means that the disk will not develop significant eccentricity, as the Kozai resonance within the individual annuli will be suppressed$^{20}$ (see SI). Recall also that here, we are ignoring dissipative effects that would generally act to circularize the disk and maintain its rigidity. Assuming that the angular momentum of the stellar binary greatly exceeds that of the proto-planetary disk, the maximal inclination that can be excited between the stellar rotation axis and the disk's orbital plane is approximately equal to twice the inclination of the proto-planetary disk with respect to the binary orbit: $\psi_{max} \simeq 2 i'$. As a result, retrograde planetary orbits can be naturally achieved, provided that the inclination of the stellar companion exceeds $45$ degrees. Figure 2 shows the time-evolution of two such examples. 

\begin{figure}
\includegraphics[width=1\columnwidth]{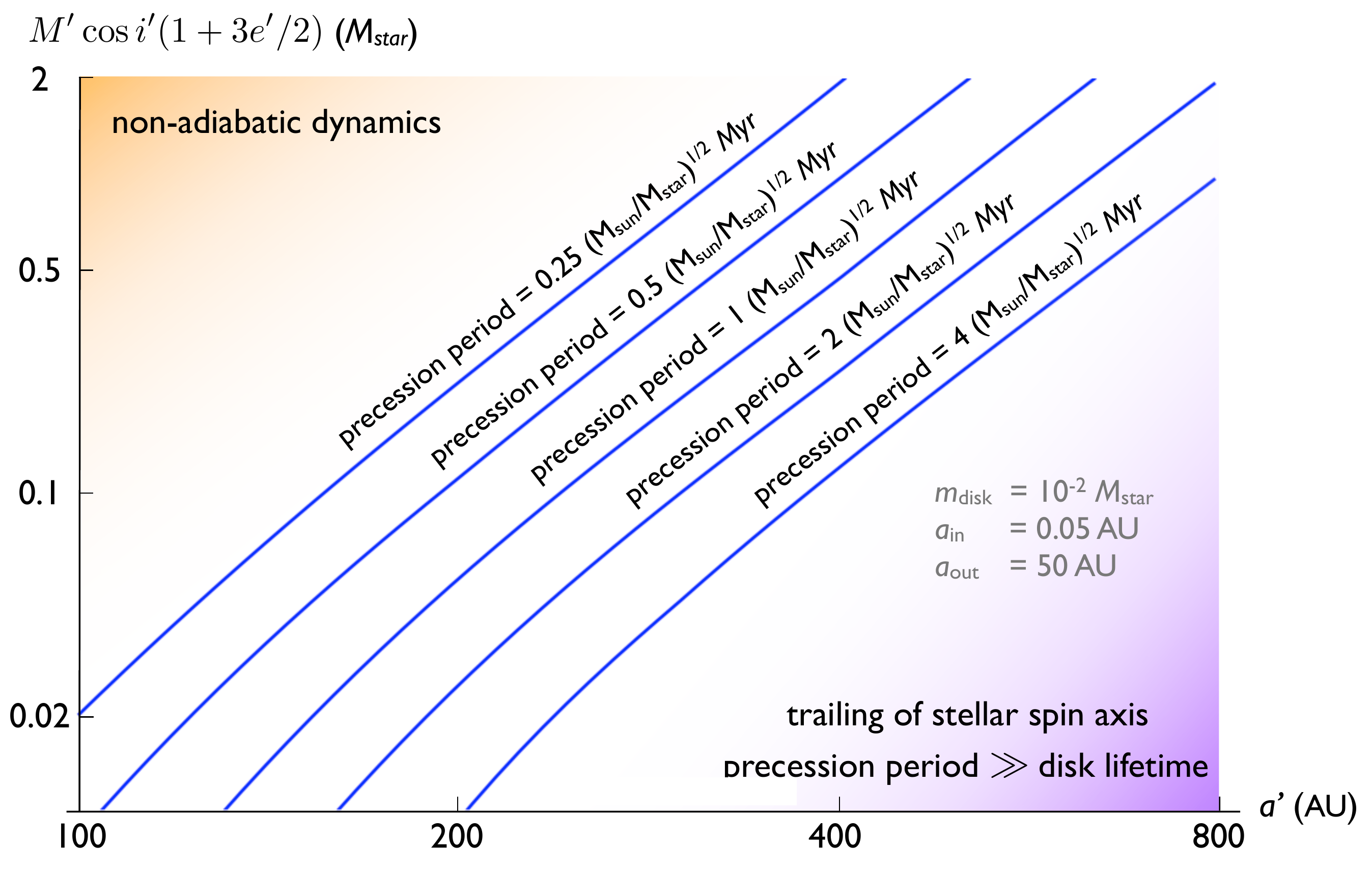}
\caption{\small{Timescales for excitation of spin-orbit misalignment. The characteristic nodal recession period of a $m_{disk} = 10^{-2} M_{star}$ disk, as a function of binary mass and orbital properties is shown. The considered disk has an outer edge at $a_{out} = 50$AU. Increasing $a_{out}$ will result in linear increase of the precession frequency. Note that the period is expressed as a scaling-law in the mass of the host star. In the region of parameter space where the precession period greatly exceeds the disk lifetime, only small misalignment angles between the disk and the host star can be excited. In principle, however, if the stellar companion does not get stripped away, the ascending node of the invariable plane of the formed planetary system can also recess. However, the degree to which this can affect a planet on a close-in orbit is sensitive to the particular architecture of the system. In the region of the parameter space where dynamics seizes to be adiabatic, misalignment is certainly attainable, but more quantitatively precise (magnetohydrodynamical) modeling is required for its characterization. Finally, on the extreme high-mass/small orbital separation end of parameter space, one could envision a scenario where a newly formed disk becomes severely twisted and eventually gets disrupted as a result of strong external perturbations. Driven by viscous dissipation, however, such a structure would likely re-collapse into a new protoplanetary disk, whose orbital plane will be close to the Laplace plane of the stellar binary orbit.}}
\end{figure}	

While the Gaussian averaging model employed above yields a rigorous representation of the secular evolution of the system, it is also computationally expensive. Fortunately, similar results can be obtained with a modified (arbitrary $i',e'$) Laplace-Lagrange analytical theory$^{18}$ (see SI), providing an avenue for efficient mapping of parameter space. We have quantified the precession timescale for a range of stellar companion masses as well as binary separations. The results are summarized in Figure 3. This calculation highlights an effective equivalence between distant massive perturbers and lower mass perturbers with smaller semi-major axes, since the recession period of the disk scales approximately as $\mathcal{T}_{disk} \propto a'^3/M'$. Thus, the precessional effect of a $M'=1M_{\odot}$ star orbiting at $a=10^3$AU is equivalent to the precessional effect  that arises from the protoplanetary disk and its host star orbiting a $\sim 10^5M_{\odot}$ star cluster at $a=0.25$ pc. It is further noteworthy that bound companions are not necessarily required for production of oblate disks, as impulsive perturbations from passing stars in the birth cluster will cause the inclination of the disk to execute a random-walk and in some cases can excite significant misalignment$^{27}$. Collectively, this explanation places the $\sim 7$ degree misalignment between the Sun's spin-axis and solar system's invariable plane into a more general, extrasolar context. Although, the process of planet formation in perturbed, warped disks certainly deserves further study. 

Given the diverse nature of the environments in which planetary systems may form, one would expect a wide range of characteristic precession timescales for proto-planetary disks. Although this does not necessarily imply an isotropic distribution of spin-orbit angles, it is quite possible that hot Jupiters which emerged from protoplanetary disks in multi-stellar systems, already resided on misaligned orbits at the time of nebular dispersion. If this is true, small spin-orbit angles must be in large part either a result of adiabatic trailing of the host stars or dissipative re-alignment of the system. Such a scenario appears to be supported by observations: hot Jupiters orbiting hot, massive stars tend to be misaligned while hot Jupiters orbiting less massive cooler stars tend to have small spin-orbit angles$^3$. This transition has been attributed to tidal re-alignment of cooler stars due to the increased size of their convective zones$^{28}$ and thus appears to be in good agreement with our model.

The consistency arguments presented above indicate that disk-driven migration in binary systems is a favorable origin of misaligned hot Jupiters. We can test this model as follows. Although multiple explanations exist for the origins of hot Jupiters, short-period multi-planet systems, which are typically less massive, have almost certainly undergone disk-driven migration. Theoretically, disk-driven migration tends to maintain coplanarity and near-resonances in the planetary systems$^{29}$, both of which have been spectacularly confirmed by the \textit{Kepler} mission$^{30}$. Within the context of the model proposed here, proto-planetary disks become misaligned with the spin axes of their host stars irrespectively of the masses and orbital radii of the newly formed planets. We predict that future observations of Rossiter-McLaughlin effect will reveal that systems of close-in coplanar sub-giant planets can be misaligned with respect to the spin axes of their host stars in the absence of significant dissipative processes. Similarly, the presence of distant, nearly circular resonant planet pairs in systems that host hot Jupiters on oblique orbits, would also point to disk-torquing as the likely mechanism for the origin of spin-orbit misalignment. High-precision radial velocity monitoring of transiting systems should directly test this prediction in the near future. \\

\textbf{Acknowledgments}  \\
I am thankful to Alessandro Morbidelli, Peter Golreich, Heather Knutson, Scott Tremaine, Josh Winn, Oded Aharonson and Fred Adams for numerous useful conversations and to Dave Stevenson and Greg Laughlin for carefully reading the manuscript. I am greatly indebted to Mher Kazandjian and Jihad Touma for providing me with the softened analytical Gaussian averaging algorithm utilized in this work and help with implementation. Finally, I am grateful to Dan Fabrycky and another anonymous referee for their careful examination of the paper and numerous excellent suggestions, which resulted in a substantial improvement of the manuscript.
\\

\footnotesize{
\textbf{REFERENCES} 

1. Lin, D.~N.~C., Bodenheimer, P., Richardson, D.~C.\ Orbital migration of the planetary companion of 51 Pegasi to its present location.\ Nature 380, 606-607 \ (1996)

2. McLaughlin, D.~B.\ Some results of a spectrographic study of the Algol system.\ Astrophys. J. 60, 22-31\ (1924)

3. Winn, J.~N., Fabrycky, D., Albrecht, S., Johnson, J.~A.\ Hot Stars with Hot Jupiters Have High Obliquities.\ Astrophys. J. 718, L145-L149 \ (2010)

4. Ghez, A.~M., Neugebauer, G., Matthews, K.\ The multiplicity of T Tauri stars in the star forming regions Taurus-Auriga and Ophiuchus-Scorpius: A 2.2 micron speckle imaging survey.\ Astron. J. 106, 2005-2023 \ (1993)

5. Kraus, A.~L., Ireland, M.~J., Martinache, F., Hillenbrand, L.~A.\ Mapping the Shores of the Brown Dwarf Desert. II. Multiple Star Formation in Taurus-Auriga.\ Astrophys. J. 731, 8 \ (2011)

6. Marks, M., Kroupa, P. \ Inverse dynamical population synthesis: Constraining the initial conditions of young stellar clusters by studying their binary populations.\ ArXiv e-prints arXiv:1205.1508 \ (2012)

7. H{\'e}brard, G., and 20 colleagues \ The retrograde orbit of the HAT-P-6b exoplanet.\ Astron. Astrophys. 527, L11 \ (2011)

8. Wu, Y., Murray, N.\ Planet Migration and Binary Companions: The Case of HD 80606b.\ Astrophys. J. 589, 605-614 \ (2003)

9. Fabrycky, D., Tremaine, S. \ Shrinking Binary and Planetary Orbits by Kozai Cycles with Tidal Friction.\ Astrophys. J. 669, 1298-1315 \ (2007)

10. Naoz, S., Farr, W.~M., Lithwick, Y., Rasio, F.~A., Teyssandier, J.\ Hot Jupiters from secular planet-planet interactions.\ Nature 473, 187-189 \ (2011)

11. Ford, E.~B., Rasio, F.~A.\ Origins of Eccentric Extrasolar Planets: Testing the Planet-Planet Scattering Model.\ Astrophys. J. 686, 621-636 \ (2008)

12. Nagasawa, M., Ida, S., Bessho, T.\ Formation of Hot Planets by a Combination of Planet Scattering, Tidal Circularization, and the Kozai Mechanism.\ Astrophys. J. 678, 498-508 \ (2008)

13.  Wu, Y., Lithwick, Y. \ Secular Chaos and the Production of Hot Jupiters.\ Astrophys. J. 735, 109 \ (2011)

14. Guillochon, J., Ramirez-Ruiz, E., Lin, D.\ Consequences of the Ejection and Disruption of Giant Planets.\ Astrophys. J. 732, 74 \ (2011)

15. Morbidelli, A., Crida, A.\ The dynamics of Jupiter and Saturn in the gaseous protoplanetary disk.\ Icarus 191, 158-171 \ (2007)

16. Marquis de Laplace, P.-S.\ Traite de mecanique celeste, 1 and 2.\ Paris : Crapelet ; Courcier ; Bachelier \ (1799)

17 Gauss, K.~F. \ Theoria motvs corporvm coelestivm in sectionibvs conicis solem ambientivm..\ Hambvrgi, Svmtibvs F.~Perthes et I.~H.~Besser, \ (1809) 

18. Murray, C.~D., Dermott, S.~F.\ Solar system dynamics. UK: Cambridge University Press, \ (1999)  

19. Touma, J.~R., Tremaine, S., Kazandjian, M.~V.\ Gauss's method for secular dynamics, softened.\ Mon. Not. R. Astron. Soc. 394, 1085-1108 \ (2009)

20. Batygin, K., Morbidelli, A., Tsiganis, K.\ Formation and evolution of planetary systems in presence of highly inclined stellar perturbers.\ Astron. Astrophys. 533, A7 \ (2011)

21. Morbidelli, A.\ 2002.\ Modern Celestial Mechanics : Aspects of Solar System Dynamics \ ~London: Taylor \& Francis, \ (2002)

22. Herbst, W., Bailer-Jones, C.~A.~L., Mundt, R., Meisenheimer, K., Wackermann, R.\ Stellar rotation and variability in the Orion Nebula Cluster.\ Astron. Astrophys. 396, 513-532 \ (2002)

23. Calvet, N., Brice{\~n}o, C., Hern{\'a}ndez, J., Hoyer, S., Hartmann, L., Sicilia-Aguilar, A., Megeath, S.~T., D'Alessio, P. \ Disk Evolution in the Orion OB1 Association.\ Astron. J. 129, 935-946 \ (2005)

24. Kaib, N.~A., Raymond, S.~N., Duncan, M.~J.\ 55 Cancri: A Coplanar Planetary System That is Likely Misaligned with Its Star.\ Astrophys. J. 742, L24 \ (2011)

25. Kraus, A.~L., Hillenbrand, L.~A.\ The Role of Mass and Environment in Multiple-Star Formation: A 2MASS Survey of Wide Multiplicity in Three Young Associations.\ Astrophys. J. 662, 413-430 \ (2007)

26. Adams, F.~C.\ The Birth Environment of the Solar System.\ Annual Review of Astronomy and Astrophysics 48, 47-85 \ (2010)

27. Bate, M.~R., Lodato, G., Pringle, J.~E.\ Chaotic star formation and the alignment of stellar rotation with disc and planetary orbital axes.\ Mon. Not. R. Astron. Soc. 401, 1505-1513 \ (2010)

28. Lai, D.\ Tidal Dissipation in Planet-Hosting Stars: Damping of Spin-Orbit Misalignment and Survival of Hot Jupiters.\ ArXiv e-prints arXiv:1109.4703 \ (2011)

29. Terquem, C., Papaloizou, J.~C.~B. \ Migration and the Formation of Systems of Hot Super-Earths and Neptunes.\ Astrophys. J. 654, 1110-1120 \ (2007)

30. Lissauer, J.~J., and 24 colleagues \ Architecture and Dynamics of Kepler's Candidate Multiple Transiting Planet Systems.\ Astrophys. J. Supplement Series 197, 8 \ (2011)
}

\clearpage
{\Large Supplementary Information}
\\
\\
In this work, our primary focus lies in understanding the long-term gravitational interactions between protoplanetary disks and distant inclined external perturbers that are gravitationally bound to the host-stars. We model the disk as a series of prograde-orbiting concentric, massive wires that are initially coplanar and nearly circular ($e \sim 10^{-3}$). We average over the Keplerian mean motion of the external perturber and also model it as a massive wire. In the example shown in the main text, the perturber is taken to have an eccentricity of $e' = 0.5$, and is inclined with respect to the disk by an inclination $i' = 45$ deg and  $i' = 70$ deg.

We work at two levels of accuracy: first we utilize an unrestricted (arbitrary $e, i$) numerical model based on Gaussian orbital averaging method. Second, we obtain quantitatively similar results with an analytical small-angle approximation to the Gaussian model, i.e. the Laplace-Lagrange secular theory. 

\section{Gaussian N-Ring Model}

	Qualitatively, Gauss's averaging method is rather intuitive: calculation of the phase-averaged interactions between two non-resonant orbits is equivalent to treating the two orbits as massive wires, where the line-density is inversely proportional to orbital velocity, and computing the forces they exert on each-other. Within the context of this approximation, the semi-major axes of the rings become constants of motion (see Ch.7 of ref. [18]). This method allows for a rigorous treatment of a conservative representation of the protoplanetary disk as well as the distant perturber provided that the number of such wires is large. In practice, the full Gaussian averaging procedure can be computationally expensive, severely limiting the number of rings in the model. Furthermore, in a conventional calculation, ring-ring crossing produces a discontinuity in the force calculation that requires high resolution in the integration. 
	
	Fortunately, a recent extension of Gauss's method has allowed for significant advances in the implementation of the algorithm described above$^{19}$. In particular, ref. [19] extended the averaging method to softened gravitational interactions. By extension of gravitational softening of a point-mass, which is equivalent to dispersing the mass over a Plummer sphere$^{31}$, softening the gravitational potential in an N-ring code allows for an approximate representation of a continuous disk with a limited number of discrete wires. Furthermore, the analytic averaging approach demonstrated in ref. [19] results in a tremendous speed-up of the integration. 
	
\begin{figure}[t]
\includegraphics[width=1\columnwidth]{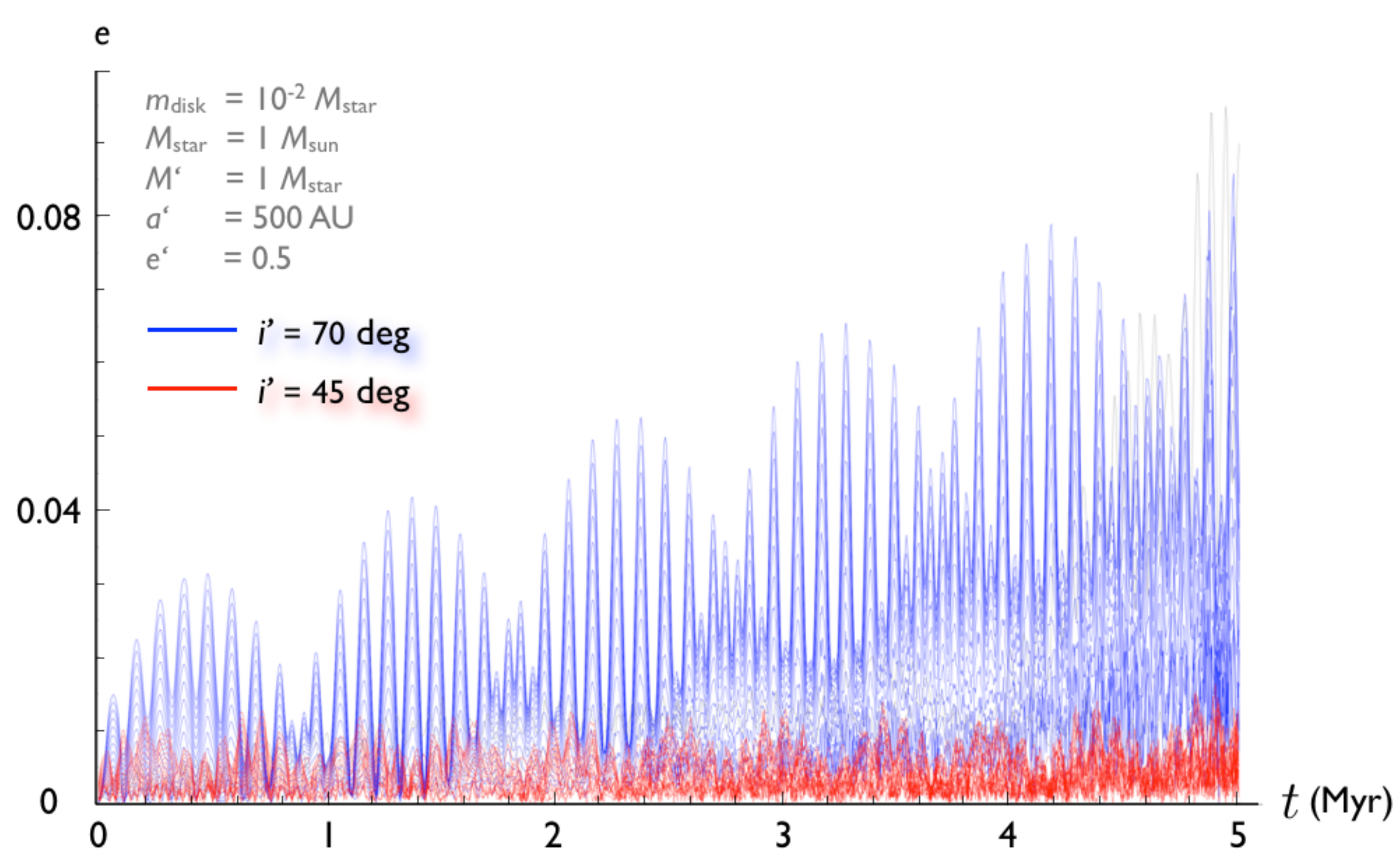}
\caption{\footnotesize{Eccentricities of the disk annuli as functions of time corresponding to the Gaussian secular solutions presented in Figure 2. Although the disk is perturbed by a massive ($M' = M_{star}$) binary companion with $e' = 0.5$, the predominantly adiabatic nature of the perturbations ensures that significant eccentricity is never excited and the Kozai resonance is erased. As in Figure 2, the red curves corresponds to the solution with $i' = 45$ deg where as the blue curves correspond to the solution with $i' = 70$ deg.}}
\end{figure}	
	
	The model employed in this work borrows directly from ref. [19]. In interest of conciseness, we shall not restate the details of the the softened Gaussian averaging method here and restrict ourselves to the particularities of our implementation. We model the disk as a system of 30 equal-mass rings, distributed in equal semi-major axis intervals (i.e. $\Sigma \propto r^{-1}$ surfrace density profile) between $a_{in} = 1$AU and $a_{out} = 50$AU, comprising $m_{disk} = 10^{-2} M_{star}$. We choose a softening length of $\epsilon = 0.79$AU (approximately half the separation between the rings). In the numerical averaging procedure, the rings are broken up into at least 128 sectors each. The equations of motion are integrated using the fourth-order Runge-Kutta method$^{32}$ with a constant timestep of $\tau = 100$ years. Increasing the number of rings by a factor of two while decreasing the softening length or decreasing $a_{in}$ did not modify the results significantly.
	
\begin{figure}[t]
\includegraphics[width=0.97\columnwidth]{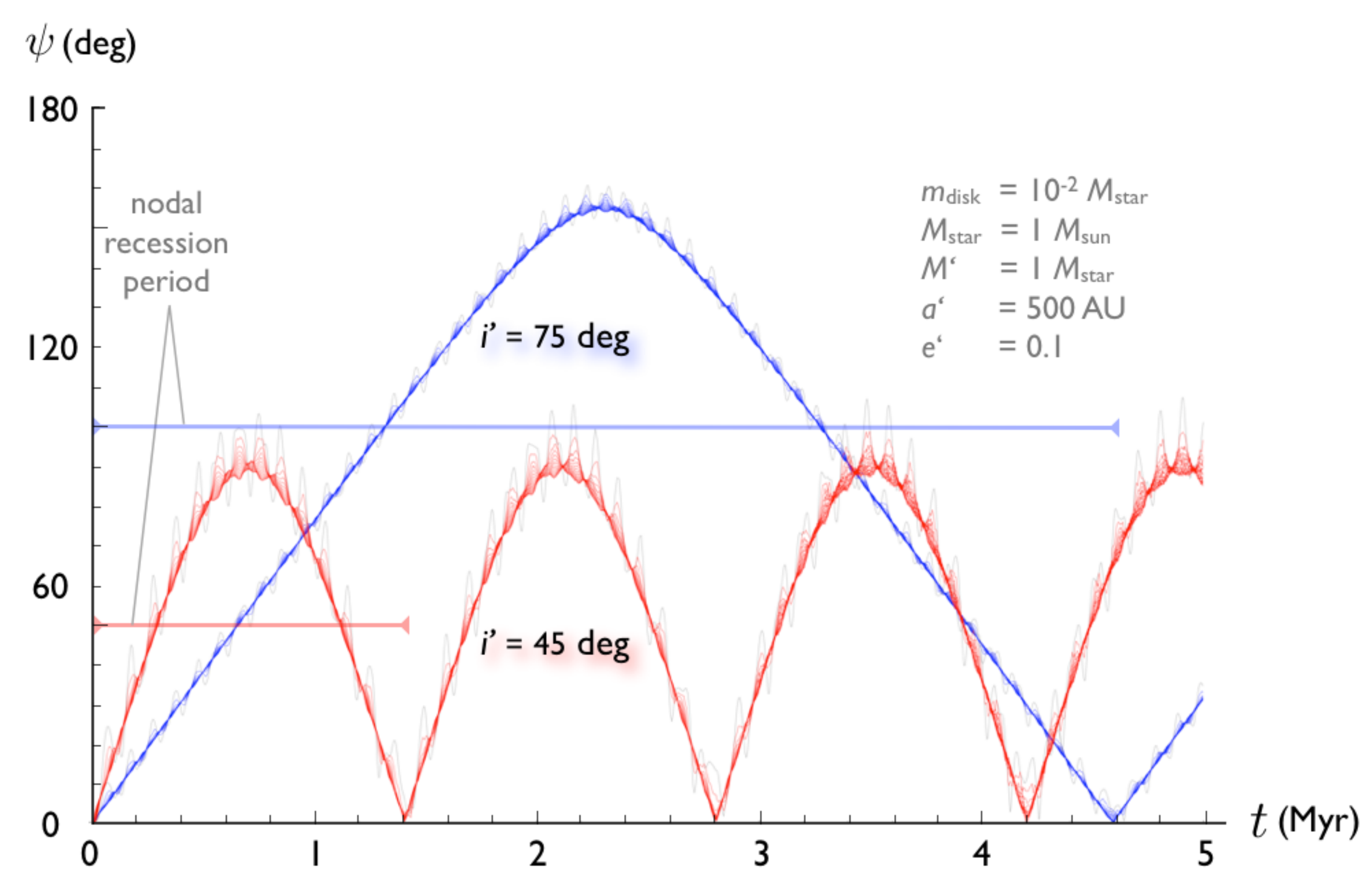}
\caption{\footnotesize{Time evolution of the stellar-spin/disk orbit normal misalignment angle. The setup is identical to that considered in the main text, with the exception of the binary companion's eccentricity (here taken to be $e' = 0.1$) and inclination (here taken to be $i'=45^{\circ}$ (red) and by $i'=75^{\circ}$ (blue)). This calculation was performed using a conservative softened Gaussian averaging model (with $N_{\rm{rings}} = 31$), and thus contains no restrictions on the secular dynamics of the system, but ignores dissipative forces of the gas. The nodal recession periods characteristic of this setup are $\mathcal{T}_{disk} \simeq 1.4$Myr ($i'=45^{\circ}$) and $\mathcal{T}_{disk} \simeq 4.6$Myr ($i'=70^{\circ}$). Throughout the duration of the integration, the annuli of the disk never attain significant eccentricity ($e \lesssim 0.01$).}}
\end{figure}
	
	The primary effect of disk self-gravity is to offset the radial frequency from the orbital mean motion, and thereby give rise to comparatively fast apsidal precession$^{33}$. This effect is of great importance to planetary formation as it acts to effectively decouple the eccentricity dynamics of the disk from that of the perturbing body. Specifically, this is accomplished as follows. To leading order, secular excitation of orbital eccentricity is governed by a term in the Hamiltonian that contains a harmonic of the form $\cos (\varpi - \varpi')$, where $\varpi$ is the longitude of perihelion and as before, the primed quantity refers to the external perturber$^{18}$. An application of Lagrange's planetary equations yields $de/dt \propto \sin (\varpi - \varpi')$. It is easy to show that significant excitation of the eccentricity can only be accomplished if $(\varpi - \varpi')$ is a slowly varying quantity (since the integrated effect of this harmonic scales inversely with $d(\varpi - \varpi')/dt$). Otherwise, the above mentioned harmonic becomes a quickly varying term, and has a small effect on the averaged evolution of the orbits (recall that the secular approximation itself is motivated by filtering out all quickly-varying effects from the Hamiltonian). In direct analogy, other higher order terms the Hamiltonian containing longitudes of perihelion will also only act to introduce low-amplitude, high-frequency noise in the eccentricity solution. In other words, because the self-interaction timescale within the disk is much shorter than the external perturbation timescale, the eccentricity dynamics of the disk predominantly exhibit adiabatic effects$^{34}$.
 
 \begin{figure}[t]
\includegraphics[width=1\columnwidth]{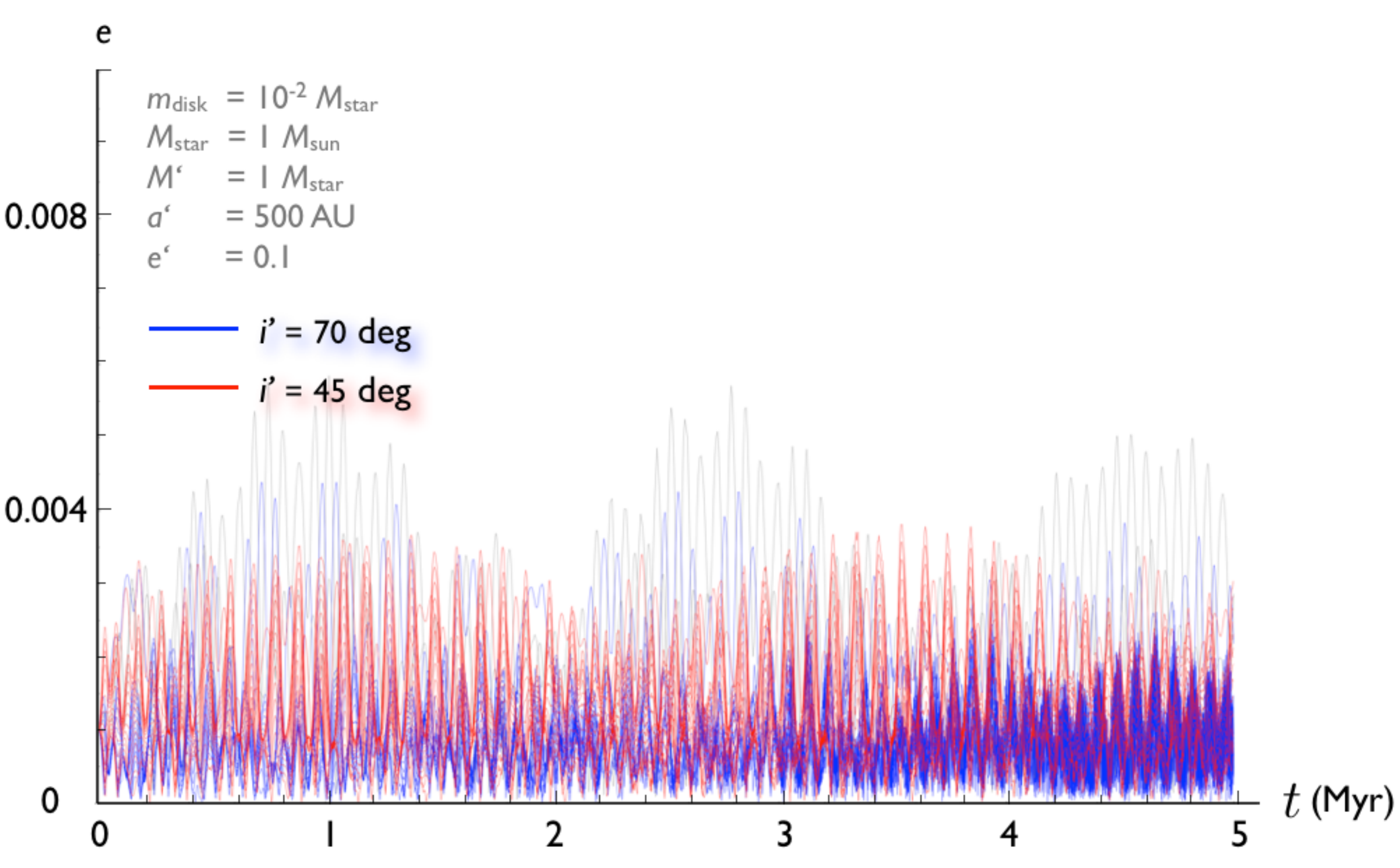}
\caption{\footnotesize{Eccentricities of the disk annuli as functions of time corresponding to the Gaussian secular solutions presented in Figure 5. See also Figure 4.}}
\end{figure}
 
	Aside from direct secular excitation of eccentricities, another seemingly detrimental effect which is erased by disk self-gravity is the Kozai resonance$^{35}$. An important property of the Kozai resonance is the libration of the argument of perihelion. If the argument of perihelion of the disk circulates quickly (as it does in the calculations presented here), the Kozai effect seizes to be a resonance and instead simply becomes another quickly-varying term in the Hamiltonian$^{20}$. These arguments are directly tested in the Gaussian N-ring model employed here, since the model explicitly contains all secular terms of the Hamiltonian. Indeed, in accord with the reasoning presented above, throughout the duration of the integrations presented in the main text, the external perturber retains $e' = 0.5$ while the eccentricities of the disk annuli never exceed $e \lesssim 0.1$ (see Figure 4). Furthermore, it is important to recall that the estimates of eccentricity excitation presented here should be viewed as an upper-limits, since dissipative forces of the gas will generally act to circularize the orbits on timescales much shorter than $\sim 1$ Myr.

	The inclination evolution of the considered disk is presented in Figure 2. An important property to be noted from the inclination solution is that all plotted annuli never depart significantly from each-other in phase-space. Instead, they always remain within $\hat{i} \lesssim $ few degrees of each-other. Physically, this means that the disk remains locally unwarped. An important consequence of this lack of warping is that standard theories of planetary formation$^{36}$ and disk-driven migration$^{37}$ are directly applicable. In other words, because the disk's ascending node is precessing sufficiently slowly for the adiabatic condition$^{34}$ to be satisfied, the processes of planetary formation and migration move forward as if the disk was completely isolated from the external perturber. 
	
	While the eccentricities within the disk remain low throughout the duration of the integrations, the ellipticity of the stellar perturber's orbit does play an appreciable role in dictating the nodal recession frequency of the disk. Specifically, a higher $e'$ corresponds to faster nodal recession (in the subsequent sections, this will be shown to be a result of the non-linear secular terms contained in the Hamiltonian). To demonstrate this effect, as well as the sensitive dependence on inclination at near-orthogonal angles numerically, we performed an additional pair of numerical experiments (setting $i' = 45$ deg and $i' = 75$ deg) with an identical setup to that described in the main text, except for a lower perturber eccentricity ($e' = 0.1$). The resulting inclination evolution is shown in Figure 5 with the corresponding eccentricity evolution shown in Figure 6. Note that although the mass and semi-major axis of the stellar perturber are held fixed, the changes in $e'$ and $i'$ make a significant difference in the nodal recession timescale compared to what is reported in Figure 2. Particularly, the recession periods characteristic of the low eccentricity ($e' = 0.1$) setup are $\mathcal{T}_{disk} \simeq 1.4$Myr ($i'=45$ deg) and $\mathcal{T}_{disk} \simeq 4.6$Myr ($i'=75$ deg), while that corresponding to the nominal ($e' = 0.5$) case are $\mathcal{T}_{disk} \simeq 0.9$Myr ($i'=45$ deg) and $\mathcal{T}_{disk} \simeq 1.8$Myr ($i'=70$ deg).

\section{Laplace-Lagrange N-Ring Model}

As already mentioned above, the single-CPU implementation of the Gaussian N-ring model is too computationally demanding to be a useful tool for mapping out parameter space. Fortunately, in the adiabatic regime where the disk remains coherent and unwarped, Laplace-Lagrange secular theory can be used to approximately reproduce the results obtained with the Gaussian model. As above, we model the proto-planetary disk as a conservative system of massive concentric rings (initialized at $i = 0$), perturbed by an inclined distant massive ring which represents the stellar companion. 

The Laplace-Lagrange secular theory is qualitatively equivalent to the Gaussian secular model employed above, however the theory explicitly assumes that all eccentricities and inclinations remain small, such that $\sin(i) \simeq i $. This assumption completely decouples the eccentricity and inclination dynamics. While neglecting the eccentricity-inclination coupling is justified for low-$e'$ perturbers, the resulting equations do not capture the dependence of the nodal recession rate on $e'$ in case the latter is significant. Thus, for the purposes of this section, we shall restrict our discussion to low-$e'$ systems and focus on the analytical reproduction of the integrations shown in Figure 5. Subsequently, we shall introduce a leading-order correction for the perturber's eccentricity in the following section.

In terms of Keplerian orbital elements, the scaled Laplace-Lagrange Hamiltonian of the j$^{th}$ ring reads:
\begin{equation}
\mathcal{H}_j = \frac{1}{2} B_{jj} i_{j}^{2} +  \sum_{j=1,j\neq{k}}^{N} B_{jk} i_{j} i_{k} \cos(\Omega_{j}-\Omega_{k})
\end{equation}
where as before, $i$ is inclination, $\Omega$ is the longitude of ascending node and $B$'s are interaction coefficients. While Keplerian orbital elements do not form a canonically conjugated set, in terms of complex Poincar$\rm{\grave{e}}$ variables $\xi = i \exp(\imath \Omega)$, the Laplace-Lagrange equations of motion form an eigen-system$^{38,39}$:
\begin{equation}
\frac{d\xi_{j}}{dt} = \sum_{k=1}^{N} \imath B_{jk} \xi_{k}.
\end{equation}
The coefficient's of the \textbf{B} matrix are:
\begin{eqnarray}
B_{jj} &=& -\frac{n_{j}}{4} \sum_{k=1, k \neq j}^{N} \frac{m_k}{M_{star}+m_{j}} \alpha_{jk} \bar{\alpha}_{jk} \tilde{b}_{3/2}^{(1)}(\alpha_{jk}) \nonumber \\
B_{jk} &=& \frac{n_{j}}{4} \frac{m_k}{M_{star}+m_{j}} \alpha_{jk} \bar{\alpha}_{jk} \tilde{b}_{3/2}^{(1)}(\alpha_{jk})
\end{eqnarray}
where $n$ denotes the mean motion, $m$ is mass, $\alpha_{jk} = a_{j}/a_{k}$ if $(a_j < a_k); a_{k}/a_{j}$ if $(a_k < a_j)$, and $\bar{\alpha}_{jk} = \alpha_{jk} $ if $(a_j < a_k)$; $1$ if $(a_k < a_j)$. In contrast to planetary secular theory, in disk secular theory it is customary to soften the Laplace coefficients, $\tilde{b}_{3/2}^{(1)}$, by the disk aspect ratio $h = H/a$ to account for its finite vertical thickness$^{40}$:
\begin{equation}
\tilde{b}_{3/2}^{(1)} (\alpha)= \frac{1}{\pi} \int_{0}^{2\pi} \frac{\cos (\phi)}{((1+\alpha^2)(1+h^2)-2 \alpha \cos (\phi))} d \phi
\end{equation}
The closed-form solution to the equations of motion reads: 
\begin{equation}
\xi_{j}(t) = \sum_{k=1}^{N} \beta_{jk} \exp{\imath(f_{k} t + \delta_k)}
\end{equation}
where $f$'s are the eigen-frequencies and $\beta$'s are eigen-vectors of the \textbf{B} matrix, and $\delta$'s are phases which are set by initial conditions$^{38}$.

\begin{figure}[t]
\includegraphics[width=1\columnwidth]{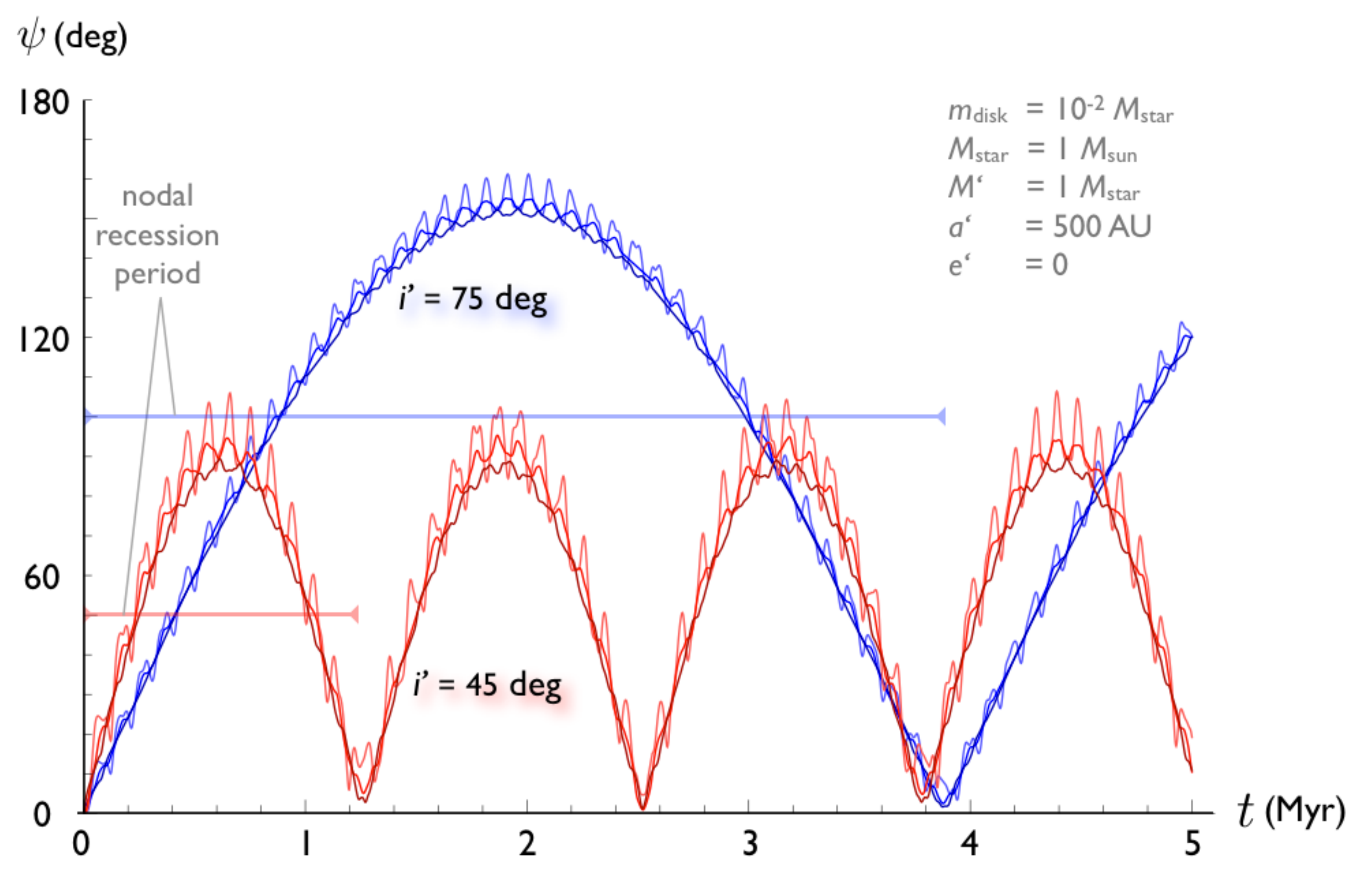}
\caption{\footnotesize{Time evolution of the stellar-spin/disk orbit normal misalignment angle, $\psi$. This solution was obtained using the Laplace-Lagrange secular theory and is aimed at reproducing Figure 5. The plotted of curves represent the annuli of the disk at $a = 1$AU, $a = 10$AU and $a = 50$AU. The nodal recession periods characteristic of this setup are $\mathcal{T}_{disk} \simeq 1.2$Myr ($i'=45^{\circ}$) and $\mathcal{T}_{disk} \simeq 4$Myr ($i'=75^{\circ}$). The agreement between the Laplace-Lagrange and Gaussian models is satisfactory, especially provided the level of approximation inherent to the Laplace-Lagrange theory (e.g. truncation of the disturbing function, $e' = 0$, etc) and a somewhat different representation of the disk.}}
\end{figure}

Because of the analytic nature of the solution, we are not as limited in N, allowing for a somewhat different representation of the disk. While, we take $a_{out} = 50$AU as above, the inner edge is considerably closer to the host star: $a_{in} = 0.05$AU. Furthermore, following ref. [41], the disk is now represented as a series of $N=100$ rings where the spacing is logarithmic with the semi-major axis ratio between neighboring annuli set to $\alpha = 0.9325$ throughout the disk. As before, the surface density profile of the disk is taken to be $\Sigma \propto r^{-1}$ and the disk mass is $m_{disk} = 10^{-2}M_{star}$.

We make one trivial modification to the standard setup of the calculation: we reduce the coefficients in the $\bf{B}$ matrix corresponding to disk-external perturber interactions by a factor of $\cos(i')$, treating only the disk's mid-plane projected component of the force exerted by the external ring as dynamically important. This introduces the $1/\cos(i')$ dependence into the forced nodal recession period of the disk approximately observed in the Gaussian secular model. 

Figure 7 shows a reproduction of the examples shown above with Laplace-Lagrange secular theory. Aside from a mild quantitative difference between the observed nodal recession periods ($\mathcal{T}_{disk}^{LL}\simeq 1.2$ Myr as compared to $\mathcal{T}_{disk}^{Gauss}\simeq 1.4$ Myr for $i' = 45$ deg and $\mathcal{T}_{disk}^{LL}\simeq 4$ Myr as compared to $\mathcal{T}_{disk}^{Gauss} \simeq 4.6$ Myr for $i' = 75$ deg) the agreement between the models is satisfactory, especially provided the level of approximation inherent to the Laplace-Lagrange theory and a somewhat different representation of the disk. 

At first glance, the applicability of classical Laplace-Lagrange secular theory in the context of the problem at hand may be surprising, given that the inclination between the disk and the external perturber is not small. However, this discrepancy is only apparent in the initial reference frame of the disk (i.e. the frame in which we measure $\psi$). In a reference frame that is centered on the host star and is normal to the stellar orbit angular momentum vector (as shown in Figure 1), the disk's inclination, $i'$, remains nearly constant in time. Rather, the quantity that changes is the disk's ascending node on which the Laplace-Lagrange theory places no restriction. It is further important to note that within the (self-gravitating) disk, the small-angle approximation poses no problem since under adiabatic perturbations considered here, the \textit{mutual} inclinations among the neighboring annuli remain small at all times. Consequently, once the effect of the large mutual inclination between the disk and the stellar orbit is taken into account by projecting the force exerted by the external perturber onto the disk's mid-plane, Laplace-Lagrange secular theory provides an adequate approximation to the adiabatic dynamical evolution of the system.

In summary, although a self-consistent account for non-secular perturbations as well as dissipative, non-gravitational effects within the disk will modify the solution quantitatively, the qualitative behavior of the system is well captured within the context of the secular models used here.

\section{An Approximate Closed-Form Solution}

As already demonstrated with the Gaussian and Laplace-Lagrange models above, protoplanetary disks remain largely unwarped thanks to their own self-gravity. We can take advantage of this fact to derive a simple, closed-form analytical solution for the secular evolution of continuous rigid disks under external perturbations. Recall that the evolution of the disk can simply be interpreted as nodal recession at constant inclination. If all annuli of the disk are fixed to its mid-plane, the disk's state vector can be represented as a single point in polar coordinates with the radius $r = |\xi| = i'$ and the polar angle $\theta = \arctan(\rm{Im}[\xi]/\rm{Re}[\xi]) = \Omega$. Accordingly, a single recession period of the disk will be represented as a circle or radius $r = i'$ centered on on the origin.

The rate of nodal recession can be calculated using Laplace-Lagrange theory as above. In particular, the disk's recession rate is given by the orbital angular momentum weighted average of the forced recession rates of the disk annuli (i.e. the diagonal components of the $\bf{B}$ matrix). Upon letting $N \rightarrow \infty$, we can turn the sum over the rings that represent the disk into an integral of surface density:
\begin{equation}
\frac{d \left< \Omega_{disk} \right> }{dt} \simeq - \frac{\int_{a_{in}}^{a_{out}} G \Sigma M' \left( r/a' \right)^2 \tilde{b}_{3/2}^{(1)} dr}{ 4 \int_{a_{in}}^{a_{out}} \Sigma \sqrt{G M_{star} r^{3}} dr} \cos (i')
\end{equation}
Similarly, the recession rate of the perturbing star's node reads:
\begin{eqnarray}
\frac{d\Omega'}{dt} &\simeq& -\frac{\pi}{2}\sqrt{\frac{G (M_{star}+M')}{a'^3}}  \cos (i') \nonumber \\
&\times& \int_{a_{in}}^{a_{out}} \frac{\Sigma}{M_{star}+M'} (r/a') \tilde{b}_{3/2}^{(1)} r dr .
\end{eqnarray}
The assumptions made up to this point generally imply that the recession rate of the perturbing body will be much slower than that of the disk itself and can be interpreted as the slow recession of the coordinate system in which the disk's state vector is measured. As a result, the period for cyclic excitation of the misalignment angle, $\psi$, is given by $\mathcal{T}_{disk} = |2 \pi/(d \left< \Omega_{disk} \right>/dt + d\Omega'/dt) |$.

\begin{figure}[t]
\includegraphics[width=1\columnwidth]{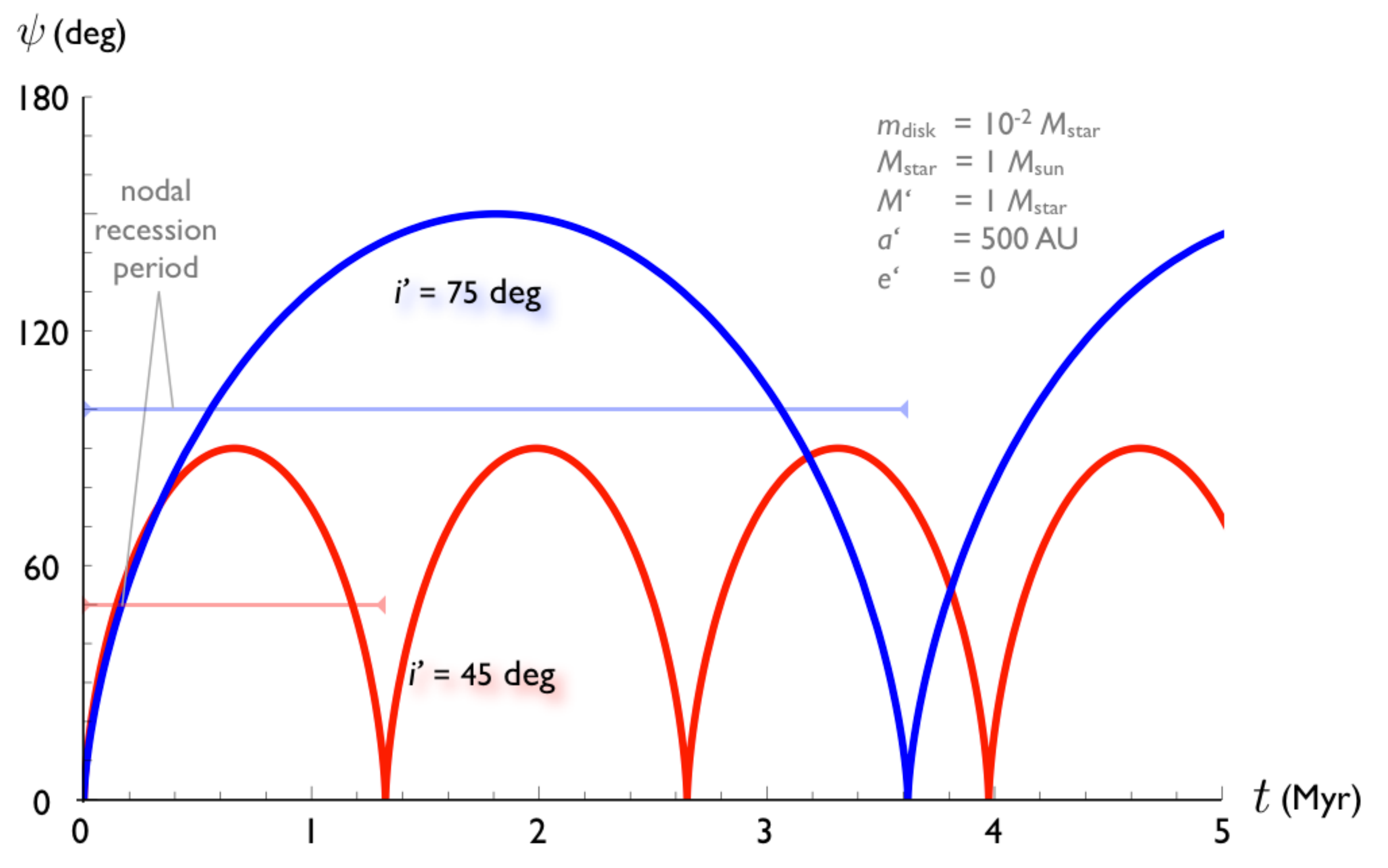}
\caption{\footnotesize{Time evolution of the stellar-spin/disk orbit normal misalignment angle, $\psi$. This solution was obtained using the approximate cycloid model, based on the Laplace-Lagrange secular theory and is aimed at reproducing Figure 5, and by extension, Figure 7 as well. The nodal recession periods characteristic of this setup are $\mathcal{T}_{disk} \simeq 1.3$Myr ($i'=45^{\circ}$) and $\mathcal{T}_{disk} \simeq 3.6$Myr ($i'=75^{\circ}$).}}
\end{figure}

\begin{figure}[h]
\includegraphics[width=1\columnwidth]{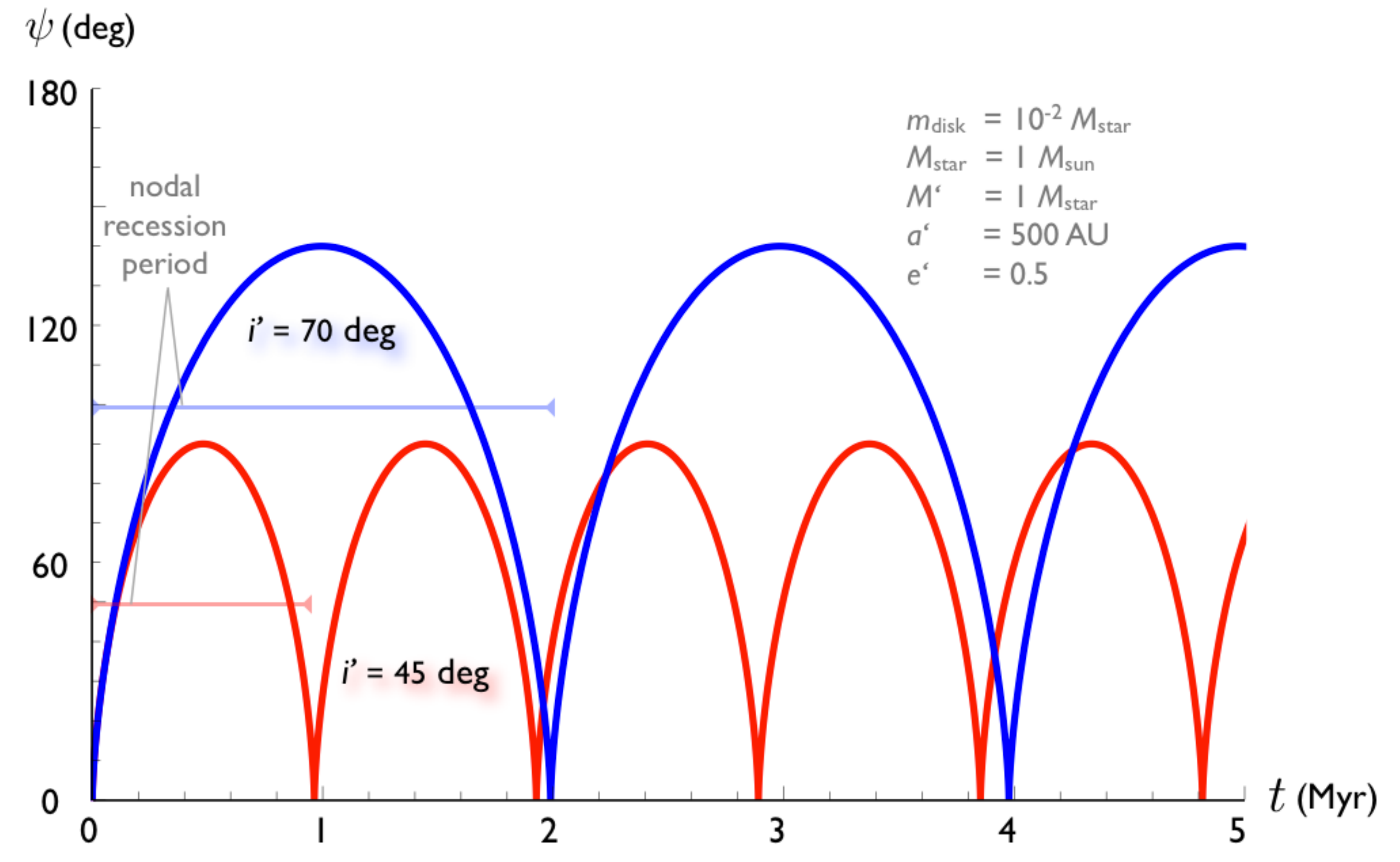}
\caption{\footnotesize{Time evolution of the stellar-spin/disk orbit normal misalignment angle, $\psi$. This solution was obtained using the approximate cycloid model, based on the Laplace-Lagrange secular theory, and corrected for the perturber's eccentricity, $e'$. The nodal recession periods characteristic of this setup are $\mathcal{T}_{disk} \simeq 1$Myr ($i'=45^{\circ}$) and $\mathcal{T}_{disk} \simeq 2$Myr ($i'=70^{\circ}$), similar to those reported in Figure 2.}}
\end{figure}

In the framework of the polar coordinates described above, the misalignment angle, $\psi$, is represented by the phase-space distance between the disk's initial condition and its state vector. As the disk's node recesses, $\psi$ will trace the rim of the disk's circular phase-space orbit and its time evolution will therefore be described by a cycloid. Specifically, the evolution of $\psi$ is given by the following parametric equation:
\begin{equation}
\{t, \psi \} = \{ \mathcal{T}_{disk} (\gamma - \sin(\gamma))/2\pi, i' (1 - \cos(\gamma))   \}
\end{equation}
where $\gamma$ is a parameter that governs the number of cycles through which the system has rotated (a single cycle corresponds to $\gamma = 2\pi$). We recalculated the evolution of $\psi$ for the example shown in Figures (5) and (7) with this simplified formalism, choosing the disk parameters as above. The calculated cycloids are presented in Figure 8. The observed nodal recession periods are quantitatively similar for low inclination to those obtained with the Gaussian model ($\mathcal{T}_{disk}^{Cy}\simeq 1.3$ Myr as compared to $\mathcal{T}_{disk}^{Gauss}\simeq 1.4$ Myr for $i' = 45$ deg and $\mathcal{T}_{disk}^{Cy}\simeq 3.6$ Myr as compared to $\mathcal{T}_{disk}^{Gauss} \simeq 4.6$ Myr for $i' = 75$ deg). It is noteworthy that the $\mathcal{T}_{disk} \propto \cos(i)$ dependence is explicitly built into the equations. Furthermore, note that although the formalism is based on the Laplace-Lagrange theory, the characteristic periods of the cycloids differ somewhat from the Laplace-Lagrange N-ring solution due to the continuous representation of the disk.

In our discussion of the analytical models based on the Laplace-Lagrange theory so far, we have neglected the effects of the coupling between the disk's inclination and the perturber's eccentricity. As already stated above, this assumption is only satisfied when $e'$ is small (a setup not necessarily characteristic of star-forming environments). Consequently, here we will extend the model developed above to incorporate a correction for the perturber's eccentricity. To leading order, the coupling between the disk's inclination and the perturber's eccentricity manifests itself as a fourth order secular effect. Accordingly, we write the new Hamiltonian of the disk as:
\begin{eqnarray}
\mathcal{H}_j &=& \frac{1}{2} B_{jj} i_{j}^{2} +  \sum_{j=1,j\neq{k}}^{N} B_{jk} i_{j} i_{k} \cos(\Omega_{j}-\Omega_{k}) \nonumber \\
 &+& \frac{1}{2} C_{jk'} i_{j}^{2} e'^{2}
\end{eqnarray}
where the newly introduced interaction coefficient reads
\begin{eqnarray}
C_{jk'} &=& - \frac{n_j}{4} \frac{m'}{M_{star}+m_{j}} \alpha_{jk'} \bar{\alpha}_{jk'} \nonumber \\
& \times & \large( 2 \tilde{b}_{3/2}^{(1)}(\alpha_{jk'}) + 4 \alpha_{jk'} \frac{\partial}{\partial \alpha_{jk'}} \tilde{b}_{3/2}^{(1)} (\alpha_{jk'}) \nonumber \\
&+& \alpha_{jk'}^2 \frac{\partial^2}{\partial \alpha_{jk'}^2} \tilde{b}_{3/2}^{(1)}(\alpha_{jk'}) )
\end{eqnarray}
and $k'$ refers to the index of the ring that represents the perturber. Generally, the fourth order secular Hamiltonian does not yield equations of motion that have analytical solutions. However, in the special case where most of the angular momentum of the system is contained in the orbit of the stellar perturber (as is the case here), the perturbers orbital elements remain nearly constant throughout the disk's lifetime. Consequently, $e'$ can be treated as a \textit{parameter} rather than a variable, rendering the newly added coupling, an effectively second order term. 

Further simplifications can be made by taking advantage of the fact that $\alpha_{jk'} \ll 1$. Specifically, we expand the Laplace coefficient as a hypergeometric series, leading to the approximations $\tilde{b}_{3/2}^{(1)} \simeq 3 \alpha$; $\alpha \ \partial/ \partial \alpha \ \tilde{b}_{3/2}^{(1)} \simeq 3 \alpha$; $\alpha^2 \ \partial^2/ \partial \alpha^2 \ \tilde{b}_{3/2}^{(1)} \simeq 135 \alpha^3/4$. Retaining only the linear terms, we find that $B_{jj}$ and $C_{jk'}$ take on similar forms. Consequently, after some rearrangement, we can rewrite the diagonal coefficients of the \textbf{B} matrix as 
\begin{equation}
B'_{jj} \rightarrow B_{jj} \left(1 + \frac{3}{2}e'^2 \right)
\end{equation}
and effectively reduce the Hamiltonian (9) to an expression that is formally identical to that of the Hamiltonian (1). This modification has a simple translation to the cycloid model introduced in this section: we can account for the perturber eccentricity by simply enhancing $d \left< \Omega_{disk} \right> /dt$ as given in equation (6) by a factor of ($1 + 3 e'^{2}/2 $). 

Using the newly modified cycloid model, we can now aim to reproduce the results shown in FIgure (2) of the main text quantitatively by explicitly setting $e' = 0.5$. The results are shown in Figure (9). The observed nodal recession periods match those obtained with the Gaussian model quite well: ($\mathcal{T}_{disk}^{Cy}\simeq 1$ Myr as compared to $\mathcal{T}_{disk}^{Gauss}\simeq 0.9$ Myr for $i' = 45$ deg and $\mathcal{T}_{disk}^{Cy}\simeq 2$ Myr as compared to $\mathcal{T}_{disk}^{Gauss} \simeq 1.8$ Myr for $i' = 70$ deg). Collectively, this analysis suggests that significant eccentricity of the perturber can diminish the timescale for excitation of spin-orbit misalignment by a factor of $\sim 2$.

\section{Adiabatic Trailing of the Host Star}

The angular momentum of a host-star of mass $M_{star}$, radius $R_{star}$, and rotation rate $\omega$ can be represented as an orbiting ring with semi-major axis
\begin{equation}
\tilde{a} = \left(\frac{16 \omega^2 k_2^2 R_{star}^6}{9 I^2 G M_{star}} \right)^{1/3}
\end{equation}
and mass
\begin{equation}
\tilde{m} = \left( \frac{3 M_{star}^2 \omega^2 R_{star}^3 I^4}{4 G k_2 }    \right)^{1/3}
\end{equation}
where $k_2$ is the apsidal motion constant, and $I$ is the stellar moment of inertia. Upon making this approximation, we can interpret the forced nodal recession period of this ring (stellar bulge) as the characteristic angular momentum coupling timescale between the disk and the stellar spin axis, $\mathcal{T}_{star} \simeq 2 \pi/\tilde{B}$. As in the previous section, we consider a continuous disk by making $N \rightarrow \infty$ and take advantage of the fact that $\tilde{a} \ll a_{in}$, again leading to the approximation $\tilde{b}_{3/2}^{(1)} \simeq 3 \alpha$. The resulting expression reads:
\begin{equation}
\mathcal{T}_{star} \simeq \frac{4}{3} \left[ \sqrt{\frac{G M_{star}}{\tilde{a}^3}} \int_{a_{in}}^{a_{out}} \frac{\Sigma}{M_{star}} \left( \frac{\tilde{a}}{r} \right)^3   r  dr \right]^{-1}
\end{equation}

We consider a typical pre-main-sequence star with $M = 1 M_{\odot}$, $R_{star} = 2R_{\odot}$, $k_{2} = 0.014$, $I = 0.08$, surrounded by the proto-planetary disk with an inner edge, $a_{in}$, at the stellar co-rotation radius and outer edge at $a_{out} = 50$ AU, comprising $m_{disk} = 10^{-2} M_{star}$ with a $\Sigma \propto r^{-1}$ surface density profile as above. This setup yields $\mathcal{T}_{star} \sim 10$ Myr and $\mathcal{T}_{star} \sim 0.3$ Myr for slow and fast rotators respectively. These estimates suggest that the spin axes of slow rotators will generally not adiabatically follow, bur rather slowly precess around their disk's angular momentum vectors. If $\mathcal{T}_{disk} \ll \mathcal{T}_{star}$, then the host-star's spin-axis will effectively remain stationary (as assumed in Figure 2 and Figure 5). However, if $\mathcal{T}_{disk} \sim \mathcal{T}_{star}$, significant excitation of mutual inclination between the disk and the host star can still take place, but the time-evolution of $\psi$ will be more complicated than the cyclic patterns presented in this work.

\end{document}